\documentclass[pra, footnotes,aps, showpacs, twocolumn]{revtex4}
\usepackage{amsmath}
\usepackage{graphicx}
\usepackage{txfonts}
\newcommand{\ket} [1] {\vert #1 \rangle}
\newcommand{\bra} [1] {\langle #1 \vert}

\begin{document}

\title{Temporal Leggett-Garg-Bell inequalities for sequential  multi-time actions in quantum information processing, and a re-definition of Macroscopic Realism }

\author{Marek \.Zukowski}
\affiliation{Institute of Theoretical Physics and Astrophysics, University of Gda\'nsk,
ul. Wita Stwosza 57, PL-80-952 Gda\'nsk, Poland\\ and Institut fuer Experimentalphysik, Universitaet Wien, Boltzmanngasse 5, A-1090 Vienna,
Austria}
\begin{abstract}
The usual formulation   of Macrorealism is recast to make this notion fully concurrent with the basic ideas behind classical physics. The assumption of non-invasiveness of measurements is dropped. Instead, it is assumed that the current state of the system defines full initial conditions for its subsequent evolution.
An example of a new  family of temporal Bell inequalities is derived which can be applied to processes in which the state of the system undergoes arbitrarily many transformations (which was not the case in the original approach). An  exponential (in terms of number of operations) violation of this inequality is demonstrated theoretically. Finally it is shown that such inequalities were indirectly tested in a 2005 experiment by the Weinfurter group.
\end{abstract}


\maketitle

\section{Introduction}
What is the principal reason for faster than-classical protocols of quantum information processing?
There are many attempts to give this answer. It was recently suggested by Brukner et al.  \cite{BRUKNERVEDRAL} that the reason, or one of the reasons, might be in violation of ''temporal
Bell inequalities". Such inequalities were introduced by Leggett and Garg \cite{LEGGETTGARG}, and were aimed at the question the relation between Quantum Mechanics and Macroscopic Realism.
They formulated the principle of Macrorealism as: 
\begin{itemize}
\item
a macroscopic system with two or more distinct states will at all times be in one  of these states,
\item
it is possible, in principle, to determine the state of the system with an arbitrarily small perturbation on its subsequent dynamics (noninvasive measurability).
\end{itemize}
They considered macroscopic quantum coherence in a SQUID, and showed that effectively there is no flux ''when nobody looks".
To this end they derived what is often called ``temporal Bell inequalities".

A different version of such inequalities was introduced by Brukner et al. \cite{BRUKNERVEDRAL} (for an earlier derivation, without a direct link with the discussion of Macrorealism, see \cite{GOLSHANI}; for an extensive study see \cite{KOFLER} and \cite{KOFLERBRUKNER}). They follow basically the same technical assumptions as the original ones, however the observer is allowed to choose between various observables which he or she wants to measure at a given instant of time. The original ones allowed the observer to freely choose the time of observation, but not the observable, which was fixed throughout the process. As what will be shown below is an extension of the Brukner et al. approach, the assumptions behind it will be now presented.

The observer has a choice between two apparatus settings for each instant of time at which he or she is to make a measurement. The measurement are to be made at instants of time  $t_o$ and later at  $t_1$. The following traits of a macro-realistic theory are assumed:
\begin{itemize}
\item
In the theoretical description one is allowed to use all variables $A_m(t)$, the values of which are eigenvalues of the observable $\hat{A}_m$, which represent the values which could have been obtained,  had the given observable been measured at time $t$, regardless what was the actual measurement. The observer has a choice $m=1,2$, or even larger. All $A_m(t)$'s are treated as unknown, but nevertheless fixed numbers, all of them at an equal footing, that is  for example the sum $A_1(t)+ A_2(t)$ has a definite, but unknown, value. (This is an assumption of {\em realism} - it is satisfied by classical systems. Please note that, if $\hat{A}_1$ and $\hat{A}_2$ are quantum observables, which do not commute, then at a given instant of time only one observable can be measured,  and thus one deals here with counterfactual statements.)
\item
{\em Non-invasiveness}: The values $A_m(t_1)$ are independent of whether or not a measurement was performed earlier,  at $t_0$, and which observable was at this earlier time measured. In short values  $A_m(t_1)$ are independent of the measurement settings chosen earlier. ({\em Note that this is a strong assumption, which does not have to hold even for classical systems when an act of observation produces a disturbance. This assumption will be modified  in the paper.})
\item 
Values $A_m(t_0)$ do not depend on what happens at later times, especially at $t_1$. 
\item  
The experimenter is free to choose the observable which is to be measured at a given instant of time. That is the choices are statistically independent of the set of values $A_m(t)$. 
\end{itemize}

With a similar type of algebra as in the case of CHSH inequalities, under the assumption that all involved eigenvalues are $\pm1$, Brukner et al. \cite{BRUKNERVEDRAL}  show that 

\begin{widetext}
\begin{equation}
E(A_1(t_0),A_1(t_1))+ E(A_1(t_0),A_2(t_1))+ E(A_2(t_0),A_1(t_1)) - E(A_2(t_0),A_2(t_1)) \leq 2,
\end{equation}
\end{widetext}
where $E(A_k(t_0),A_m(t_1))$ stands for a correlation function, understood as an averaged product of the results, that is $\langle A_k(t_0)A_m(t_1)\rangle$.
One can easily check that, in an attempt to construct multi-time temporal Bell inequalities, say a Mermin type extension
\begin{widetext}
\begin{equation}
E(A_1(t_0)A_1(t_1)A_2(t_2))+ E(A_1(t_0)A_2(t_1)A_1(t_2))+ E(A_2(t_0)A_1(t_1)A_1(t_2))- E(A_2(t_0)A_2(t_1)A_2(t_2)) \leq 2,
\end{equation}
\end{widetext}
one faces the fact that it cannot be violated more strongly than the previous one. This contrasts the case of the usual Bell multi-party inequalities, which in the GHZ case \cite{GHZ} are violated much more strongly than for two qubits, see \cite{BRUKNERVEDRAL}. 

Let me explain this feature, with an example. Take a qubit, and use its spin $\frac{1}{2}$ representation. In such a case projectors representing eigenstates of a Pauli operator $\vec{n}\cdot\vec{\sigma}$, with $|\vec{n}|=1$, are given by $\frac{1}{2}(1+v\vec{n}\cdot\vec{\sigma})$, where $v=\pm1$ is the eigenvalue. Thus if one starts with qubit in an arbitrary state $\varrho=\frac{1}{2}(1+\vec{s}\cdot\vec{\sigma})$, where $|\vec{s}|\leq 1$, and recalls that sequential quantum measurements form a Markov process, then the correlation function for  measurements with the Stern-Gerlach directions $\vec{a}$, $ \vec{b}$ and finally $\vec{c}$ reads
\begin{equation}
E(\vec{a}, \vec{b}, \vec{c})=\sum_{klm=\pm1}klm P(k,l,m)=(\vec{s}\cdot\vec{a})(\vec{b}\cdot\vec{c}),
\end{equation}
where $P(k,l,m)$ denotes the probability of a sequence of results, $k,l,m$ is a consecutive order. Note that this correlation function factorizes, while the one for a sequence of two measurements
\begin{equation}
E(\vec{a}, \vec{b})=\sum_{kl=\pm1}kl P(k,l)=\vec{a}\cdot\vec{b},
\end{equation} 
does not, and what is crucial here, is formally identical (up to a sign) with the usual  correlation function for two qubits in a singlet state.  Note that if the initial state is pure noise, $|\vec{s}|=0$, 
the three measurements correlation function vanishes.

The same problems arise when one considers the original Leggett-Garg inequalities. In the Heisenberg picture different moments of observation lead to different observables, as $\hat{A}(t)= U^{\dagger}(t,t_0)\hat{A}(t_0)U(t,t_0)$, where $U(t,t_0)$ is the unitary evolution operator. 

One intuitively feels that there must exist some form of temporal Bell inequalities that are applicable to arbitrarily long quantum processes, which involve many instants of time, at which the system changes its state due to an external intervention. Below, such a family of inequalities will be presented. An entirely new approach will be taken, which surprisingly uses softer, more physically justified, assumptions concerning Macrorealism than the one presented above. The term Macrorealism will be still used, as the whole idea will be illustrated with something that resembles a quantum informational protocol. In such a case one is tempted to compare qubits, on which certain operations are performed, with a changing state of a some sort of microprocessor element (a transistor), the set of states of which represents the values of a bit (current - no current).

\section{New inequalities}
Imagine a microprocessor element  which can be in two states. The states will be denoted as $A$. For the sake of an easier mathematical representation, we shall assume that $A=\pm1$, that is the bit value $b$ represented by the state is related with respect to $A$ in the following one-to-one way $A=(-1)^{b}$.  Assume that at each instant of time $t_k$, where $k=0,1,2,...$ and $t_k<t_{k+1}$,  an operation is performed on the system which may change the value of $A$. The operation is governed by two external input bits. For the given moment they are represented by two random numbers $x_k,y'_k$, and the pair will be denoted $X_k$, and at certain points we shall assume that $X_k$ has a numerical value $2^1{y'_k}+2^0 x_k=0,1,2,3$. We assume that each $y'_k$ is completely random, whereas the distribution of $x_k$'s may be governed by a probability distribution $p(x_1, x_2,...)$. For technical reasons we assume that $x_k=0$ or $1$ and we replace $y'_k$ by $y_k=(-1)^{y'_k}$. Thus, $y_k=\pm 1$. 
After say $l$ operations the current state of the system is denoted as $A_n=A(X_1,X_2,..., X_l)$. However, we shall assume that the system forgets the reason why it is in the current state, that is the state after the $k$-th instant of time is given by
\begin{equation} 
A_m(t_k)=F_m(X_k, A_{m_{k-1}}(t_{k-1})), \label{NEWCONDITION}
\end{equation}
that is defined by the state if the system before the last operation,  $A_{m_{k-1}}(t_{k-1})$ , and by the last operation, defined by $X_k$ (this seems quite sensible in the case of classical operations on computer elements).  $F_{m}$ denotes a binary function.

We shall demand that the operations performed on the system are aimed to give at the end of the process $A_n$ which is an answer to the question about the value of the task function 
\begin{equation}
T_n=\prod_{l=1}^ny_l\cos\big(\frac{ \pi }{2}\sum_{k=1}^n x_k\big), \label{TASK}
\end{equation} 
under the promise that the distribution of $x_k$'s obeys the following probability $p(x_1, x_2,...)=2^{-N+1}|\cos(\frac{ \pi }{2}\sum_{k=1}^n x_k)|$. This simply implies that the bits $x_k$ are promised to  satisfy always the following constraint: $(\sum_{k=1}^{N} x_k){\textrm{mod} 2} = 0$, that is, are distributed in such a way that their sum is always even.  Note, that under such a promise $T_n=\pm1$. What is the average chance to get a correct result for systems obeying the above assumptions? This will be given here by the average of the product of the answer with the correct value: $\langle A_nT_n \rangle_{avg}$, where the average is over all possible values for $X_k$'s. Obviously only if this average equals $1$ the answer is always correct. If it is zero, he answer is random, uncorrelated with $T_n$

Of course, the above story does not have to be taken literally. We shall now derive an inequality which is obeyed by the average value of $A_n=A(X_1,X_2,..., X_n)$, under the restrictions given above, especially (\ref{NEWCONDITION}). From the technical point of view the derivation is resembles the case of communication complexity problems studied in \cite{HARALD}, however the interpretation of the process is different. Please note, that this was also the case in the standard approach discussed  in the introduction. One has a different interpretation of the symbols involved in the temporal inequalities, however the derivation of the actual bounds  follows the same mathematical steps as in  the case of standard Bell inequalities.

Let us write first explicitly the expression the maximum of which we search for:
\begin{widetext}
\begin{equation}
\langle A_nT_n \rangle_{avg}=\sum_{x_1,x_2,..., x_n=0,1}\sum_{y_1,y_2,...,y_n=\pm 1}\frac{1}{2^n}p(x_1,..., x_n)A(X_1,X_2,..., X_n)\prod_{l=1}^ny_lf(x_1,...,x_n), \label{FIDELITY}
\end{equation}
\end{widetext}
where $ f(x_1,...,x_n)=\cos(\frac{ \pi }{2}\sum_{k=1}^nx_k)$. 
Note that  $A_n=F_n(x_n, y_n, A_{n-1})$, and that it is a binary function of its  three arguments. It must depend on  $A_{n-1}$ because only $A_{n-1}$ might contain information about $y_{n-1}, y_{n-2},..., y_1$, which is absolutely necessary for an attempt to get the correct value of $T_n$. Please, look at equation  (\ref{TASK}): all $y_l$'s must be known in order to get the correct value. There are very few binary functions of a binary variable, just four. Let us use this fact. Treat  $x_n$ and $y_n$ as fixed, thus we have $A_n=B_{x_n,y_n}(A_{n-1})$. Because it is binary, it can only have the following form: $$B_{x_n,y_n}(A_{n-1})= D_{x_n,y_n}+C_{x_n,y_n}A_{n-1},$$ where both $C$ and $D$ are equal $\pm1$ or $0$, and $C_{x_n,y_n}D_{x_n,y_n}$=0. If $C\neq0$ then it must be of the form $C_{x_n,y_n}=c_n(x_n)y_n$, the same holds for $D$, that is one must have $D_{x_n,y_n}=d_n(x_n)y_n $. This because a term is not proportional to $y_k$ gives a vanishing input into (\ref{FIDELITY}), as for an arbitrary $g(x_n)$ one has
\begin{equation}
\sum_{y_n=\pm1}y_n g (x_n)=0.
\end{equation}
Thus $B_{x_n,y_n}(A_{n-1})= d(x_n)y_n+c(x_n)y_nA_{n-1}$. However upon one more summation over $y_{n-1}$ one has 
\begin{equation}
\sum_{y_n, y_{n-1}=\pm1}y_ny_{n-1}( d(x_n)y_n+c(x_n)y_nA_{n-1})=\sum_{ y_{n-1}=\pm1}y_{n-1}c(x_n)A_{n-1}.
\end{equation}
As we see the optimal form of $A_n=F_N(X_n, A_{n-1})$ is $y_nc(x_n)A_{n-1}$. With a similar step one shows that the optimal form of $A_{n-1}$ is  $y_{n-1}c(x_{n-1})A_{n-2}$, and so on. Continuing like that we arrive at the final formula which is 
\begin{equation}
\langle A_nT_n \rangle_{avg}=\sum_{x_1,x_2,..., x_n=0,1}K(x_1,..., x_n)\prod_{k=1}^nc_k(x_k), \label{FIDELITY2}
\end{equation}
where all $c_k(x_k)$ take values $\pm 1$, and the coefficients $K$ are given by $K(x_1,..., x_n)=p(x_1,..., x_n)f(x_1,..., x_n)$. This is mathematically isomorphic with a multi-party Bell inequality, and its bound is given by  $$\sum_{x_1,x_2,..., x_n=0,1}K(x_1,..., x_n)\prod_{k=1}^nc_k(x_k)\leq 2^{-N+1},$$
 where N=n/2 for $n $ even and $N=\frac{n+1}{2}$ for $n$ odd. As a matter of fact for $n=3$ one has a structure which is equivalent to the Mermin inequality, and the whole set is equivalent to the series of inequalities derived by Mermin in 1990 \cite{MERMIN}. 
Similar series of Bell-like inequalities were derived for the communication complexity problems in  \cite{HARALD} (see also \cite{CC-BELL}).

Note that we have just established that for $n$ odd one has
\begin{widetext}
\begin{equation}
\sum_{x_1,x_2,..., x_n=0,1}\sum_{y_1,y_2,...,y_n} \prod_{l=1}^ny_l\cos\big(\frac{ \pi }{2}\sum_{k=1}^nx_k\big)\prod_{k=1}^nc_k(x_k) A(X_1,X_2,..., X_n)
\leq 2^{\frac{n-1}{2}} 2^{n}, \label{INEQUALITY2}
\end{equation} 
\end{widetext}
where we have factored out the trivial part of the bound,  $2^n$, which is due to the $y_k$'s.

\section{Macrorealism: new formulation}
Please note that this a temporal Bell inequality, which is applicable to a system which undergoes a series of transformations governed by external parameters $X_k$.
The following modified Macrorealism is behind it:
\begin{itemize}
\item
{\em Realism}:  In the theoretical description one is allowed to use all variables $A_m(t)$, the values of which are eigenvalues of  observables $\hat{A}_m$, which represent the value which could be obtained if the given observable were measured at time $t_k$. The observer has a choice $m=X_k$ (in our example  $X_k$ can take four values). All $A_m(t_k)$'s are treated as unknown, but nevertheless fixed numbers, all of them at an equal footing, that is  for example, for two different input values, $X_k $ and $X'_k$, the expressions like  $A_{X_k}(t)\pm A_{X'_k}(t)$ have a definite, but perhaps  unknown, value. (This the old assumption, slightly rewritten to fit the studied case.)
\item
{\em Classical causality}: The values $A_m(t_{k+1})$ are not {\em directly} dependent on operations which were performed earlier, at $t_{k}$. However,  values  $A_m(t_{k+1})$  might depend on the earlier ones, that is on  $A_{m_k}(t_{k})$, which are defined by  the state of the system after the previous operation $m_k$ at $t_k$. I stress once more, there is   no direct dependence on the operation done earlier. ({\em Note that this is a an assumption which holds for the states of transistors in microchips. In classical mechanics it is  equivalent to a statement that we do not care what was the reason for the current state of an object, we care only about the state. We do not need to know {\em why} a classical particle has this or that momentum and this or that position at the given moment. Still these values are full  initial conditions for further dynamics. All systems, which follow Hamilton dynamics, including classical fields,  satisfy this condition.}) 
\item 
{\em Causality}: Values $A_m(t_{k-1})$ do not depend on what happens at later times, especially at $t_k$. (Unchanged.)
\item  
{\em Freedom}: The experimenter is free to choose the operation which is to be to be performed at a given instant of time. That is the choices are statistically independent of the set of values $A_m(t)$. (Unchanged.)
\end{itemize}

Note that these assumptions are quite general, and apply to observables endowed with any eigenvalues. When applied to our example, they are isomorphic with the set stated at the beginning of the derivation of the inequality, and the tacit assumptions used during the derivation (esp., freedom). Information theoretic inequalities involving many measurements, were introduced earlier by Morikoshi \cite{MORIKOSHI}. However they follow a completely different approach and were based on the old definition of Marcrorealism. As it is suggested in \cite{CC-BELL} one can derive inequalities involving different task functions and promises, related to e.g. the Bell inequalities discovered in \cite{WZ} and \cite{WERNERWOLF}.

\section{The quantum protocol}

This inequality is violated by a process which was experimentally realized by the group of Weinfurter \cite{HARALD}. 
In the ideal
quantum version of the protocol  one starts 
 with a qubit in the state
$\ket{\psi_i}=2^{-1/2}(\ket{0}+\ket{1})$. Then one acts
sequentially  on the qubit with the unitary phase-shift
transformation of the form $ \ket{0}\bra{0} +
e^{i\pi/2X_k}\ket{1}\bra{1}$, in accordance with the local inputs $x_k,y_k$.
After all $N$ phase shifts the state is
\begin{equation}
\ket{\psi_f} = \frac{1}{\sqrt{2}}(\ket{0}+e^{i\pi/2(\sum_{k=1}^{n}
X_k)}\ket{1}).
\end{equation}
Due to the constraint that the sum over all $X_k$ must be even (see the derivation of the inequality),
the phase factor $e^{i\pi/2(\sum_{k=1}^{n} X_k)}$ is equal to the
dichotomic function $T_n$ to be computed. Therefore, a measurement
of the qubit in the basis given by  $ 2^{-1/2}(\ket{0}+\ket{1})$ and 
$2^{-1/2}(\ket{0}-\ket{1})$ reveals the value of $T_n$, with
fidelity $ \langle A_nT_n \rangle_{avg}= 1$.
Note, that this implies that inequality (\ref{INEQUALITY2}) is violated exponentially (in terms of the number of operations $n$).

\section{Conclusions}
These findings can be generalized in many obvious ways.
Note that the prime moral of the story is that we cannot steer the state of a transistor, by sequential operations, each  governed by pairs of bits $x_k, y_k$, following a certain promise, so that at the end of the process it would give the proper value of $T_n$, given by (\ref{TASK}). In contrast, this can be easily done with a qubit. With perfect accuracy. {\em Of course, the presented inequality is just a first example of the infinitely many that can be derived using the principles presented in this work. These do not have to be constrained to two-state systems, and the inputs can be even continuous} (for a ready example, compare the communication complexity problems in \cite{HARALD} and \cite{CC-BELL}). The basic requirement is that the Macrorealistic system under consideration has a finite information capacity \cite{BP}.

\section{Acknowledgments}
This work is a part of the Q-ESSENCE project (VII FP EU). Author thanks {\v C}aslav Brukner, Johannes Kofler,  Marcin Markiewicz and Marcin Paw{\l}owski for discussions and remarks on the manuscript.

\end{document}